\documentclass[12pt]{iopart}

\usepackage{citesort}
\usepackage{cite}
\usepackage{iopams}  
\usepackage{amstext}
\usepackage{indentfirst} 
\usepackage{graphicx}
\usepackage{subfigure}
\usepackage{mathrsfs}
\usepackage{varioref}
\usepackage[applemac]{inputenc}

\graphicspath{{Figures/}}

\begin{document}

\title[On the optimal density profile for quasi-monoenergetic beams]{Ion acceleration in electrostatic collisionless shock: on the optimal density profile for quasi-monoenergetic beams}

\author{E. Boella$^1$\footnote{Present address: Mathematics Department, KU Leuven, Leuven, Belgium}, F. Fi\'uza$^2$, A. Stockem Novo$^3$, R. Fonseca$^{1,4}$, L.~O. Silva$^1$}
\address{$^1$ GoLP/Instituto de Plasmas e Fus\~ao Nuclear, Instituto Superior T\'ecnico, Universidade de Lisboa, Lisbon, Portugal}
\address{$^2$ SLAC National Accelerator Laboratory, Menlo Park (CA), USA}
\address{$^3$ Institut f{\"u}r Theoretische Physik, Lehrstuhl IV: Weltraum- und Astrophysik, Ruhr-Universit\"{a}t Bochum, Bochum, Germany}
\address{$^4$ DCTI/ISCTE, Instituto Universitario de Lisboa, Lisbon, Portugal}

\ead{elisabetta.boella@ist.utl.pt}

\vspace{10pt}
\begin{indented}
\item[]{\today}
\end{indented}

\begin{abstract}
A numerical study on ion acceleration in electrostatic shock waves is presented, with the aim of determining the best plasma configuration to achieve quasi-monoenergetic ion beams in laser-driven systems. It was recently shown that tailored near-critical density plasmas characterized by a long-scale decreasing rear density profile lead to beams with low energy spread [F. Fiúza \textit{et al.}, Physical Review Letters \textbf{109}, 215001 (2012)]. In this work, a detailed parameter scan investigating different plasma scale lengths is carried out. As result, the optimal plasma spatial scale length that allows for minimizing the energy spread while ensuring a significant reflection of ions by the shock is identified. Furthermore, a new configuration where the required profile has been obtained by coupling micro layers of different densities is proposed. Results show that this new engineered approach is a valid alternative, guaranteeing a low energy spread with a higher level of controllability.
\end{abstract}

%
\vspace{2pc}

%
%
%
%

\section{Introduction} \label{intro}

Laser pulses incident on plasma targets are capable of exciting strong accelerating fields that allow the acceleration of ions to high energies on short distances. This is why there has been a lot of interest in the topic of laser-driven ion acceleration over the past twenty years. Such a compact and affordable ion source has many potential applications in science and medicine, which have been limited until now by the cost, the size and the technological issues connected to conventional accelerator devices \cite{Malka-NP-2008, Macchi-RMP-2013}. However, despite of these advantages, there are still several difficulties that need to be addressed and overcome (such as increasing the particle energy, improving the spectral and angular control of the beam, the conversion efficiency from the laser energy into the ion beam, the stability of the acceleration parameters, etc.) before laser driven ion acceleration can be considered a mature technology. This has motivated a significant theoretical, numerical and experimental effort devoted to understanding and optimizing the physics behind the process of ion acceleration.

Shock wave acceleration has been suggested as a mechanism of ion acceleration by Denavit in 1992 \cite{Denavit-PRL-1992} and Silva \textit{et al}. in 2004 \cite{Silva-PRL-2004}: the electric field associated with the electrostatic shock can act indeed as a ``moving wall" reflecting and accelerating the plasma ions during its propagation. In these previous works, the conditions under which shocks are generated in solid targets were studied: the use of an ultra-intense laser ($I \ge 10^{20} \,$W/cm$^2$) is required to achieve efficient electron heating and the narrow ion energy spectrum is smeared out by a strong Target Normal Sheath Acceleration (TNSA) field building up at the back of the target due to expanding electrons \cite{Macchi-RMP-2013}. 

Recently, electrostatic shocks have been identified as the physical phe\-nom\-e\-non responsible for the acceleration of ions in the interaction of intense laser pulses with hydrogen gas jets \cite{Haberberger-NP-2012, Tresca-PRL-2015, Antici-exploded, Chen-Marija}. Further theoretical and numerical investigations showed the importance of using tailored near-critical density plasmas in order to get better control over the charge separation field, which develops at the back of the gas target and decreases the beam quality  \cite{Fiuza-PRL-2012, Fiuza-PoP-2013}. In particular, it has been demonstrated that an exponentially decreasing plasma profile, which leads to  a constant accelerating sheath field \cite{Grismayer-PoP-2006}, results in the acceleration of ions with a quasi monoenergetic spectrum. In the present work, the requirements on the spatial scale of the rear plasma derived in Fi\'uza \textit{et al.} \cite{Fiuza-PRL-2012} are reviewed and tested with numerical simulations. In order to focus on the physics of the sheath field, a more simplified configuration is used, where shock waves are driven by sharp density discontinuities in plasmas composed of hot electrons and cold ions, akin to laser-generated plasma systems. Moreover, a valid alternative constituted by several micro layers of decreasing density is proposed and investigated in simulations. Such an engineered approach has already been tested in the context of TNSA acceleration \cite{Nakamura-PoP-2010, Sgattoni-PRE-2012}, but for the first time, it is explored in the context of shock acceleration.

\section{On the optimal scale length to generate monoenergetic ions} \label{s:exp_plasma}

\begin{figure}[]
\begin{center}
\includegraphics[scale=0.3]{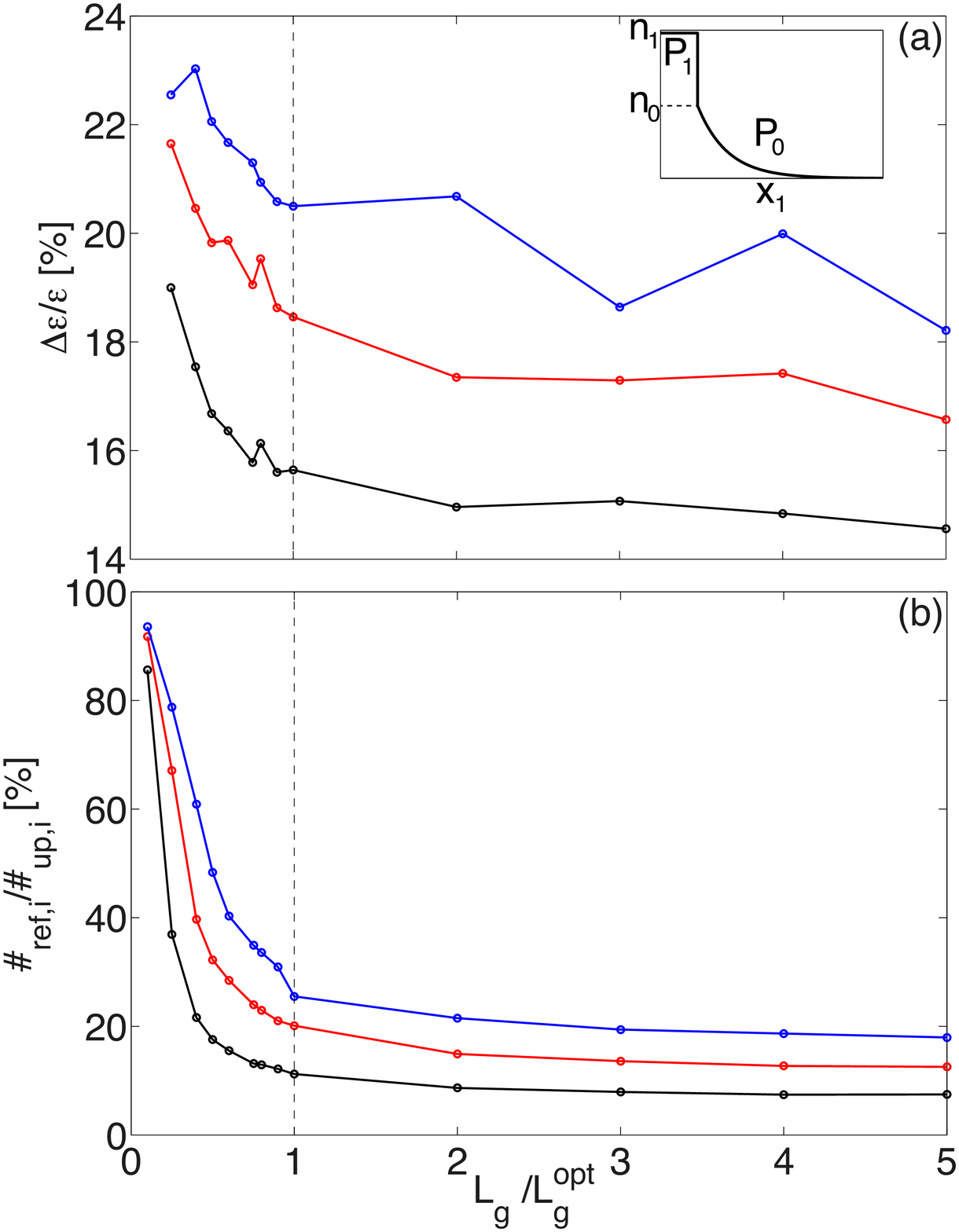}
\end{center}
\vspace{-15pt}
\caption[Upstream ion energy spread and percentage of reflected upstream ions versus $L_g$ at different times]{Upstream ion energy spread (a) and percentage of reflected upstream ions (b) versus $L_g$ at $t=3308$ (black), 4734 (red) and 6688 (blue) $\omega_{pe,1}^{-1}$, where $\omega_{pe,1}=\sqrt{4 \pi e^2 n_1/m_e}$ is the electron frequency of the plasma $P_1$. Plasma slabs with $n_1 = 10^{19} \, \text{cm}^{-3}$, $L_1 = 100 \, \mu \text{m}$, $\Gamma = 10$, $L_g^{opt} = 200 \, \mu \text{m}$, initial electron temperature $T_e = 0.08 \, \text{MeV}$ and initial density profile as shown in the inset in (a) have been considered. The parameter scan has been performed with the shell algorithm.}\label{Lg_time_ref_pmax}
\end{figure} 

It is well known that short intense laser pulses incident on solid density targets with micrometers thicknesses generate MeV electrons that, after propagating through the target, leave it at the rear surface setting up an electrostatic field $E_{TNSA}$. In the case of an abrupt plasma-vacuum transition, the field has a maximum amplitude \cite{Mora-PRL-2003} of $\sqrt{2}k_BT_e/e\lambda_D$, where $k_B$ is the Boltzmann constant, $T_e$ the electron temperature, $e$ the elementary charge and $\lambda_D=\sqrt{k_BT_e/4\pi e^2 n}$ is the Debye length for a plasma of density $n$. However, if the density decreases exponentially with a scale length $L_g$, as in the case of a controlled target pre-expansion, the electric field generated by the escaping relativistic electrons is constant at early times and its amplitude \cite{Mora-PoP-2005} is $E_{TNSA}=k_BT_e/eL_g$. 

In the case of laser-driven shock wave acceleration in near critical density target, the laser not only contributes to the shock formation inside the target, but also provides a strong heating, causing the electrons to leave the target and set up a non-uniform charge separation field, which then accelerates the ions on the rear side of the plasma to different velocities $v_{0i}$. When the strong shock propagating from inside the target outward at a constant velocity will reach this region of plasma, it will then pick up the ions accelerated by the TNSA and reflect them to a velocity~\cite{Fiuza-PoP-2013}
\begin{equation}
v_i \simeq 2Mc_{s0}+v_{0i}, \label{ref_ion_velocity}
\end{equation}
where $M>1$ is the shock Mach number and $c_{s0}=\sqrt{k_B T_e/m_i}$ is the sound speed for ions with mass $m_i$, so that the shock velocity is given by $v_s=Mc_{s0}$ \cite{Fiuza-PoP-2013}. Controlling the sheath field in a way that $v_{0i}$ is as homogeneous as possible is then crucial for obtaining a beam with low energy spread and, therefore, a uniform TNSA field, as the one due to an exponential plasma profile, is a necessary condition.  An optimal value for the scale length $L_g$ can then be determined considering the role of the competing accelerating fields.

In order to get a beam with low energy spread, it is important to guarantee that the velocity of the expanding upstream ions $v_{0i}$ is small compared to the shock velocity $v_s$, so that the shock can pick them up at the time $\tau_r$ when the shock is formed. Considering a finite density gradient, the upstream ion velocity can be expressed as  \cite{Grismayer-PoP-2006}:
\begin{equation}
v_{0i}=c_{s0}^2 \tau_r/L_g \label{prova1}.
\end{equation}
In the case of strong shocks ion reflection is the dominant dissipation mechanism and therefore the ion reflection time is similar to the shock formation time.
A numerical estimate of this quantity is given by Forslund and Shonk \cite{Forslund-PRL-1970} in the case of low Mach number shocks
\begin{equation}
\tau_r=\frac{4\pi}{\omega_{pi}}, \label{ref_time_fors}
\end{equation}
where $\omega_{pi}=\sqrt{4 \pi e^2 n/m_i}$ is the downstream ion plasma frequency. Combining equations \eref{prova1} and \eref{ref_time_fors} and imposing $v_s \gg v_{0i}$, a constraint for monoenergetic ions is obtained \cite{Fiuza-PRL-2012}:
\begin{equation}
L_g \gg \frac{4\pi c_{s0}^2}{v_s\omega_{pi}} . \label{lower_limit_Lg}
\end{equation}

\begin{figure}[]
\begin{center}
\includegraphics[scale=0.3]{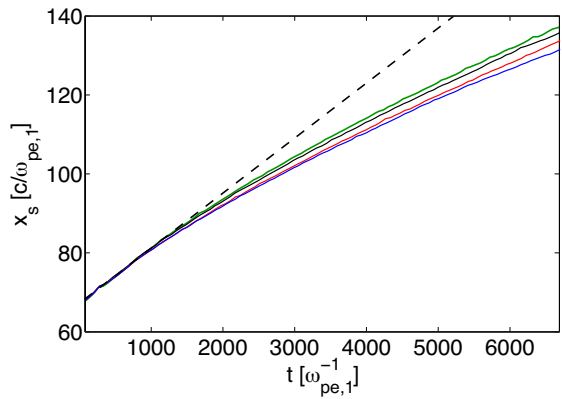}
\end{center}
\vspace{-15pt}
\caption[Shock speed versus time for different $L_g$]{Shock position $x_s$ versus time for $L_g=0.5 \, L_g^{opt}$ (green), $L_g^{opt}$ (black), $5 \, L_g^{opt}$ (red), and  $20 \, L_g^{opt}$ (blue) for the same plasma parameters considered in figure \ref{Lg_time_ref_pmax}. Simulations have been performed with the shell algorithm.}\label{shock_speed_no_const}
\end{figure}

Besides this condition for monoenergetic ion generation, there is a stricter requirement  that concerns the ideal target thickness for optimal plasma heating. Numerical simulations of laser driven shocks show that a uniform heating is a key factor to generate a stable nonlinear wave with constant velocity \cite{Fiuza-PRL-2012}. This condition imposes an upper limit on $L_g$:
\begin{equation}
L_g \le \frac{\pi c}{\omega_{pi}} , \label{upper_limit_Lg}
\end{equation}
where $c$ is the velocity of light in vacuum.

\begin{figure}[]
\begin{center}
\includegraphics[scale=0.3]{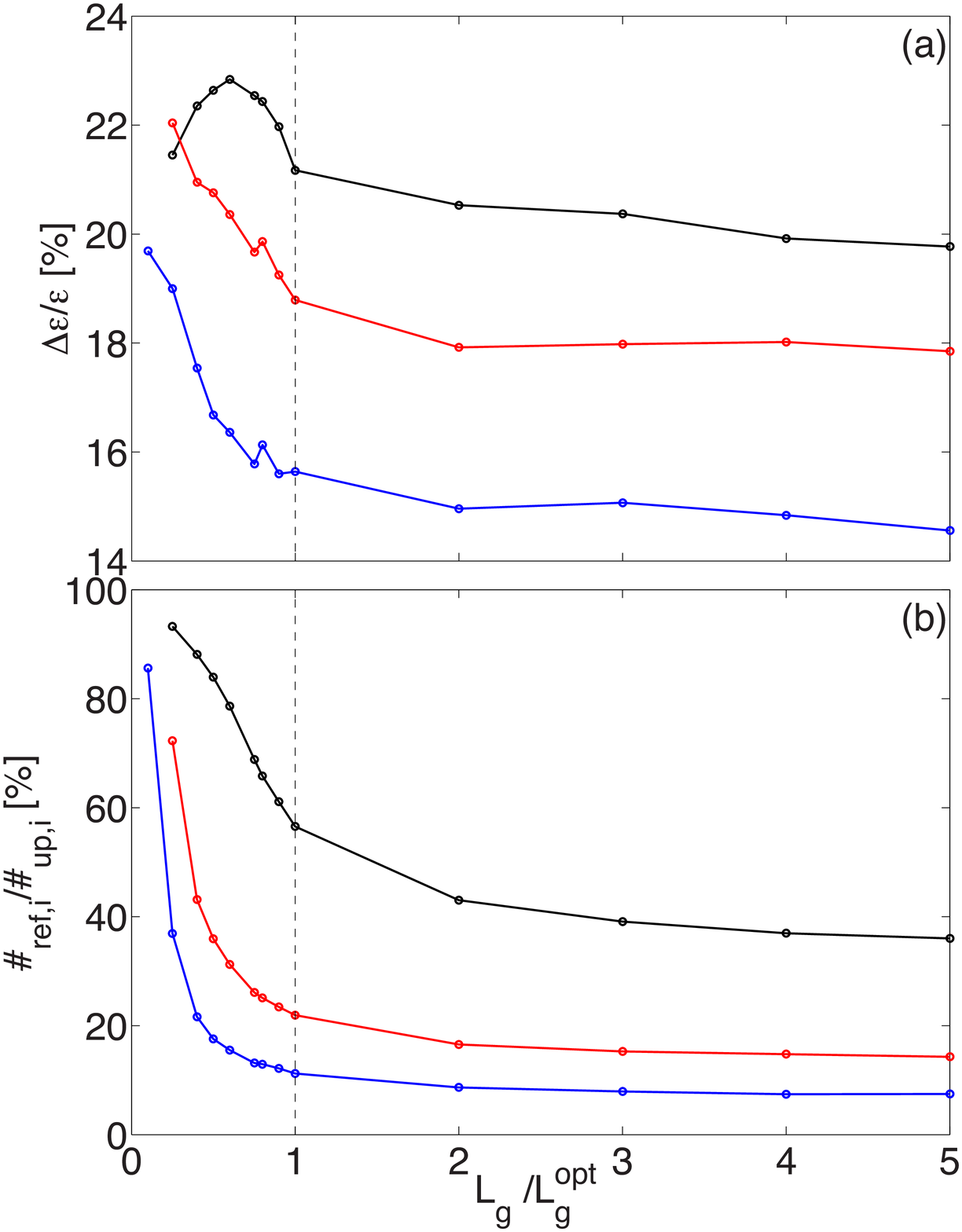}
\end{center}
\vspace{-15pt}
\caption[Upstream ion energy spread and percentage of reflected upstream ions versus $L_g$ for different initial electron temperatures]{Upstream ion energy spread (a) and percentage of reflected upstream ions (b) versus $L_g$ at $t=3308 \, \omega_{pe,1}^{-1}$ for $T_e=0.08$ (blue), $0.2$ (red) and $0.5$ (black) MeV. Plasma slabs with $n_1 = 10^{19} \, \text{cm}^{-3}$, $L_1 = 100 \, \mu \text{m}$, $\Gamma = 10$ and $L_g^{opt} = 200 \, \mu \text{m}$ have been considered. Simulations have been performed using the shell algorithm.}\label{Lg_temperature_deltaeps_ref}
\end{figure}

Assuming the upper limit of inequality \eref{upper_limit_Lg} as the optimal scale length $L_g$, theoretical predictions have been tested with simulations. The numerical study has been performed employing the shell model \cite{Dangola-PoP-2014, Boella-JPP-2016, Boella-CPC-2017}, a particle-based gridless algorithm. The use of this technique, which has been benchmarked with the particle-in-cell code Osiris \cite{Fonseca-LNCS-2002}, showing excellent agreement \cite{APS-Poster}, results particularly advantageous in this case, because it entails a reduced computational time.

The interaction between two contiguous plasma slabs $P_0$ and $P_1$ composed by cold protons and hot electrons having temperature $T_e$ has been modeled. While the slab $P_1$ has a constant density $n_1$, the slab $P_0$ has a longitudinal density profile described by 
\begin{equation}
n_0(x)= \frac{n_1}{\Gamma}\exp \left (- \frac{x - L_1}{L_g} \right) \quad \text{for} \quad x>L_1, \label{exp_density_profile}
\end{equation}
where $\Gamma = n_1/n_0(x=L_1) >1$ and $L_1$ is the length of the plasma $P_1$ (see the inset in figure \ref{Lg_time_ref_pmax} for a sketch of the initial plasma density profile). A detailed parameter scan has been performed varying $L_g$ around its optimal value $L_g^{opt} = \pi c/\omega_{pi}$, where $\omega_{pi} \simeq \omega_{pi,1}$ with $\omega_{pi,1}$ the ion plasma frequency of the slab $P_1$ has been assumed. It is important to underline that in these sets of simulations, the electron distribution is initialized as Maxwellian with uniform temperature in the box, therefore condition \eref{upper_limit_Lg} does not directly apply, however, since it imposes an upper limit on $L_g$ for laser-plasma systems, we do consider this requirements in defining $L_g^{opt}$. Figure \ref{Lg_time_ref_pmax} shows the energy spread of the reflected ion beam  and the percentage of reflected upstream ions for different values of $L_g$ at different times. All the curves present a trend $\propto L_g^{-1}$ and flatten around $L_g \simeq L_g^{opt}$, confirming the validity of the theoretical prediction. The inverse proportionality of the energy spread and the percentage of reflected ions respect to $L_g$ is strictly interconnected and can be interpreted within the theoretical  framework for ion acoustic shocks when including the reflection of ions by the electrostatic potential \cite{Stockem-PRE-2013, Cairns-POP-2014, Cairns-PPCF-2015}. Considering Maxwellian upstream ions, the number of reflected ions by the shock depends on the thermal spread of the upstream ions, besides the magnitude of potential barrier, the shock speed and the electron temperature. All other conditions being equal, the number of reflected ions will be greater for higher upstream ion velocity spread. In the present simulations, the velocity spread of these ions is determined by the spatial variation of the charge separation field which develops at the plasma rear and which in turns depends on $L_g$. In particular, the shorter the density gradient, the larger the velocity spread reached by the upstream ions prior the arrival of the shock, which leads then to a higher number of reflected ions. However, since these ions have a non-uniform velocity distribution, they will be reflected by the shock to different velocities according to equation \eref{ref_ion_velocity}, thus resulting in an ion beam with a larger energy spread. We observe that the beam energy spread increases with time. In fact, the shock speed, constant at early times, decreases at later times. This is due to the fact that the wave is constantly transferring energy to the ions and since no plasma is injected into the simulation, it slows down as a consequence of the dissipation, as illustrated in figure \ref{shock_speed_no_const} and previously noted by Macchi \textit{et al.} \cite{Macchi-PRE-2012}. The speed of the reflected ions is then no longer constant and a chirp is introduced into the ion spectrum, which causes the energy spread to increase. Furthermore the deceleration of the shock wave depends on the scale length $L_g$. Right after the shock is formed, it starts to move with the same speed, independent of the decaying scale length. However, when ion reflection becomes important, at around $t=1000 \, \omega_{pe,1}^{-1}$, it starts to lose energy and it decelerates with a rate increasing with $L_g$, in agreement with the model for soliton-like laser pulses propagating in inhomogeneus plasmas \cite{Tsintsadze-PRE-1998}. Figure \ref{Lg_temperature_deltaeps_ref} shows the energy spread of the reflected ion beam and the percentage of reflected upstream ions versus $L_g$ for different initial values of electron temperature at $t= 3308 \, \omega_{pe,1}^{-1}$. Also in this case, all the curves show the same trend and suggest that the upper limit \eref{upper_limit_Lg} on $L_g$, critical in laser driven ion acceleration, leads to the lowest energy spread, since the minimum value of $\Delta \varepsilon/\varepsilon$ is reached around $L_g=L_g^{opt}$. Moreover, as predicted by the theory, numerical simulations indicate that the optimal decay length does not depend on the initial electron temperature.
\begin{figure}[]
\begin{center}
\includegraphics[scale=0.3]{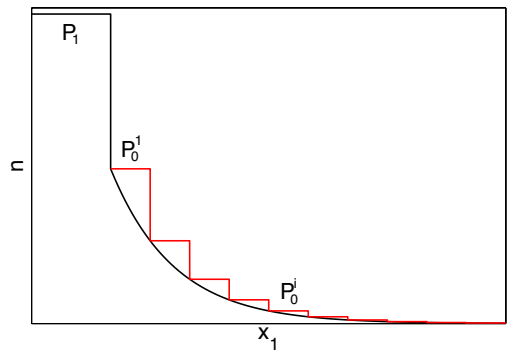}
\end{center}
\vspace{-15pt} 
\caption[Multilayer plasma (sketch)]{Sketch of the multilayer plasma. The exponentially decaying plasma profile \eref{exp_density_profile} has been replaced by several plasma layers to approximate it.}\label{multilayer}
\end{figure}

\section{Shock formation and ion reflection in multilayer plasmas}

\begin{figure*}[]
\begin{center}
\includegraphics[scale=0.3]{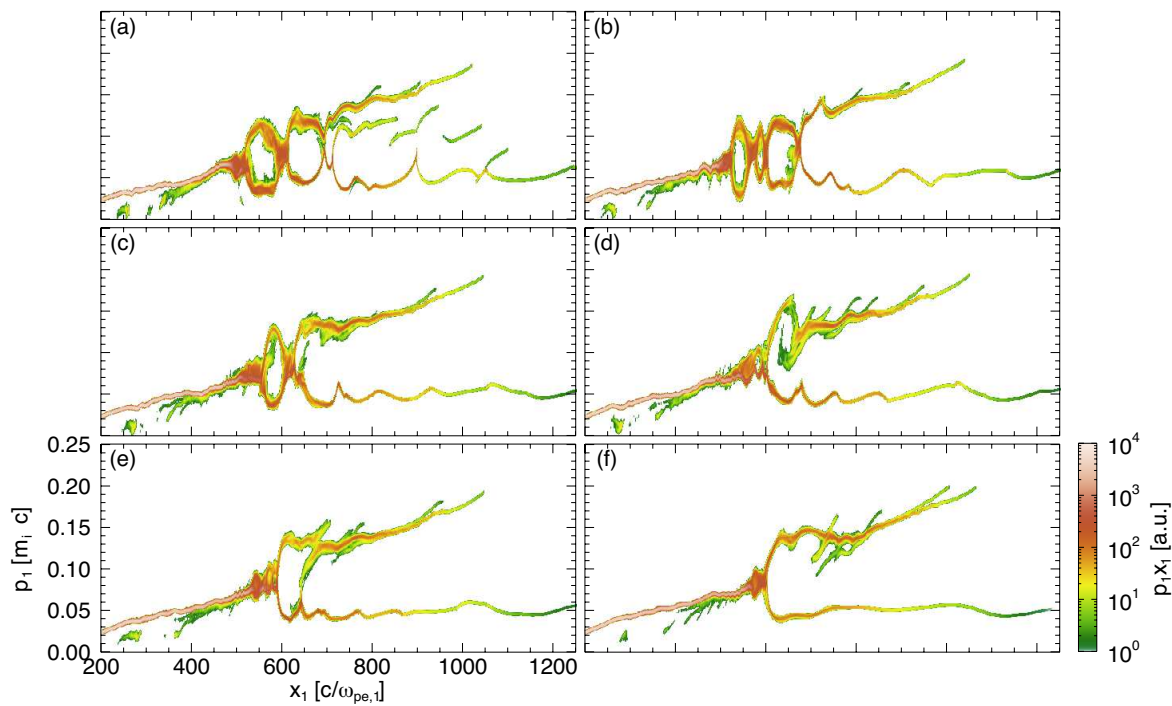}
\end{center}
\vspace{-15pt}
\caption[Multi-layer plasmas: ion phase space]{Ion phase space generated in the interaction between a plasma slab $P_1$ and a plasma $P_0$ composed of several layers with thickneses $L_0^i=350$ (a), 250 (b), 220 (c),  180 (d) and 150 (e) $\mu$m. (f) Ion phase space for density profile of $P_0$ according to equation \eref{exp_density_profile}. Simulations have been performed using Osiris.}\label{layert_target_phase_space}
\end{figure*}

A key parameter for the generation of monoenergetic ions is represented by the electric field at the interface between plasma and vacuum, which can be controlled if the upstream plasma has the required exponentially decaying density profile \cite{Fiuza-PRL-2012}. However, since an exponentially decreasing density profile with the desired features cannot be easily produced in an experiment, the possibility of using several plasma slabs with progressively decreasing density to mimic the exponential profile of equation \eref{exp_density_profile} has been tested. 

\begin{figure*}[]
\begin{center}
\includegraphics[scale=0.23]{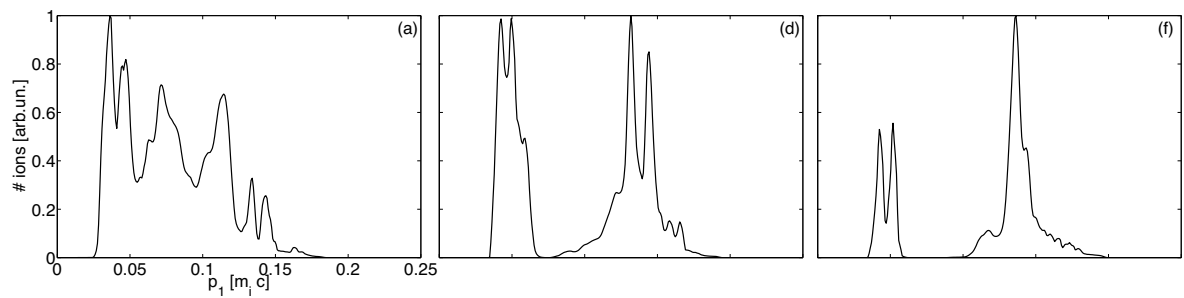}
\end{center}
\vspace{-15pt}
\caption[Multi-layer plasmas: ion distribution]{Ion momentum distributions for the cases of figure \ref{layert_target_phase_space} (a), (d) and (f)}\label{layert_target_idf}
\end{figure*}

Numerical simulations in which the plasma $P_0$ has been replaced by several plasma layers $P_0^i$  (with $i=1,...,N$, where $N$ is the total number of layers) with decreasing density (as shown in figure \ref{multilayer}) have been performed with the particle in cell code Osiris \cite{Fonseca-LNCS-2002}. The slab $P_1$ is $300 \, \mu \text{m}$ long and has a density of $n_1=10^{19} \, \text{cm}^{-3}$. The density ratio $\Gamma$ between $P_1$ and $P_0^1$ is 10. The length of the layers following $P_1$ and their density have been varied. Figure \ref{layert_target_phase_space} shows the ion phase space at $t=5600 \, \omega_{pe,1}^{-1}$ for an upstream plasma with density given by equation \eref{exp_density_profile} with $L_g=L_g^{opt}=200 \, \mu \text{m}$ and for an upstream plasma composed of several layers with lengths $L_0^i=350$, 250, 220,  180 and 150$\, \mu$m, corresponding to a density ratio $\Gamma_0^i$ between two contiguous layers $P_0^i$ of  5.4, 3.3, 2.8, 2.3 and 2 respectively. The density discontinuity triggers the generation of non linear structures at the interfaces between the layers. These structures are stronger for higher $\Gamma_0^i$ and can lead to the formation of smaller, secondary shock waves that reflect the upstream ions and give rise to ion trapping, as can be seen in figure \ref{layert_target_phase_space} (a)-(c). Their presence degrades the quality of the accelerated ion beam, leading to a spectrum which is no longer monoenergetic. For smaller values of $\Gamma_0^i$, modulations in the upstream still occur (figure \ref{layert_target_phase_space} (d) and (e)), but their influence on the property of the reflected ions is weak. Figure \ref{layert_target_idf} shows the ion distribution for the cases of figure \ref{layert_target_phase_space} (a), (d) and (f). While in the case of figure \ref{layert_target_idf} (a), the ion distribution is wide, in the case of figure \ref{layert_target_idf} (d), two peaks corresponding to the expanding upstream and to the reflected ions, can be clearly identified. The distribution is qualitatively similar to the one of figure \ref{layert_target_idf} (f) obtained with the exponentially decaying profile and the energy spread is measured of the order of 10\% in both cases. Thus, simulations show that multilayer plasmas can be a promising alternative for achieving a high quality ion beam, provided that the density discontinuity between two contiguous layers is $< 2.8$. 

\section{Summary} \label{conclusion}

Optimal conditions for ion shock wave acceleration have been presented and tested with numerical simulations in a fundamental configuration where the shock is driven by a density discontinuity between two plasma slabs. Despite of its simplicity, this framework represents well the shock formation triggered by an intense laser pulse.

A detailed parametric study of the conditions for which the optimal plasma scale length derived by Fi\'uza \textit{et al.} \cite{Fiuza-PRL-2012} is valid has been performed. Two main features of the shock accelerated ion beam have been taken under consideration: the energy spread and the fraction of the ions that has been reflected by the shock. It has been shown that for $L_g \ll L_g^{opt}$ the number of reflected ions increases (up to 90\% of the total upstream ions), as well as the energy spread, deteriorating thus the beam quality. When $L_g \simeq L_g^{opt}$, the quality of the ion beam improves, but its charge decreases. Moreover, it has been demonstrated that the value of $L_g^{opt}$ is independent of the plasma temperature.

A new configuration has been tested where the exponentially decreasing plasma profile has been replaced by several plasma layers. It has been shown that a quasi-monoenergetic ion beam ($\Delta \varepsilon / \varepsilon \simeq 10 \%$) can be achieved, provided that the density between two contiguous plasma layers decreases by a factor less than 2.8.
Such an engineering approach has the advantage of offering a high control in laser-target experiments of shock acceleration. Finally, we notice that an appropriate target suitable for this purpose could be manufactured depositing, for instance, several layers of foam at progressively decreasing densities on a solid density support using the pulsed laser deposition technique \cite{Prencipe-PPCF-2016}.

\section*{Acknowledgements}
This work was partially supported by the European Research Council (ERC-2010-AdG Grant No. 267841 and ERC-2015-AdG Grant No. 695008). Simulations were performed at the Accelerates cluster (Lisbon, Portugal) and on the supercomputer SuperMUC (Leibniz Research Center, Germany) under PRACE allocation. We would like to thank Dr. L. Fedeli (Politecnico di Milano) for fruitful discussions.

\section*{References}
\bibliography{paper_iop}

\end{document}